\documentclass[11pt]{article}
\usepackage{arxiv}
\usepackage[utf8]{inputenc}
\usepackage{amsmath,amssymb}
\usepackage{graphicx}
\usepackage{booktabs}
\usepackage{multirow}
\usepackage{xcolor}
\usepackage{hyperref}
\usepackage{cite}
\usepackage{tabularx}
\usepackage{siunitx}
\usepackage{subcaption}
\usepackage[capitalize]{cleveref}
\usepackage{tikz}
\usepackage{svg}
\usetikzlibrary{positioning,arrows.meta,fit,backgrounds,calc,decorations.pathreplacing}
\usepackage{algorithm}
\usepackage{algorithmic}
\usepackage{array}
\usepackage{tabularx}
\usepackage{booktabs}

\newcolumntype{Y}{>{\raggedright\arraybackslash}X}
\newcolumntype{Z}[1]{>{\raggedright\arraybackslash}p{#1}}

\hypersetup{colorlinks=true, linkcolor=blue, citecolor=blue, urlcolor=blue}

\newcommand{\etal}{\textit{et al.}}

\title{Approximate Attention Weighting\\
       for Sustainable FPGA-Based Vision Transformer Inference}

\usepackage{authblk}

\author[1]{Muhammad Usman}
\author[2]{Muhammad Akmal Shafique}
\author[3]{Shujaat Khan}
\author[1]{Dorit Merhof}

\affil[1]{Faculty of Informatics and Data Science, University of Regensburg, 93053 Regensburg, Germany \protect\\ \texttt{\{muhammad.usman,dorit.merhof\}@ur.de}}
\affil[2]{Department of Computer and Software Engineering, CEME, NUST, Pakistan  \protect\\ \texttt{akmal.shafique@yahoo.com}}
\affil[3]{Department of Computer Engineering, College of Computing and Mathematics, KFUPM, \protect \\Dhahran, 31261, Saudi Arabia \protect\\ \texttt{shujaat.khan@kfupm.edu.sa}}

\begin{document}
\begingroup
\renewcommand\thefootnote{}
\footnotetext{Under Review (IEEE 2026 Sustainable Energy and Industry Conference (SUSTAIN))}
\endgroup

\maketitle
\begin{abstract}
Vision Transformers have reshaped computer vision by using self-attention to capture global context across image regions. This makes them attractive for edge visual inspection and monitoring in applications such as renewable-energy infrastructure, industrial quality control, medical imaging, and autonomous-system sensing. However, deploying ViTs on small FPGAs remains challenging because the softmax stage in self-attention requires exponential evaluation and normalization, which are costly in hardware. Existing implementations often rely on CORDIC pipelines or BRAM-based look-up tables, increasing area and power consumption. This paper presents a BRAM-free approximate attention-weighting unit for FPGA-based ViT inference. The proposed design approximates the natural exponential in softmax using a 16-segment piecewise-linear function implemented entirely with distributed LUT fabric. Unlike base-2 approximations, the natural-exponential formulation preserves the pre-trained attention temperature and avoids model-specific recalibration. Implemented on a Xilinx Zynq-7020, the complete attention-row core uses 1444 LUTs, 77 DSPs, and no BRAM, while hardware-accurate emulation shows accuracy within a \(0.20\%\) absolute top-1 difference from the exact-softmax reference on ViT-family models. These results demonstrate the potential of the proposed core for energy-efficient ViT inference on resource-constrained edge-AI platforms.
\end{abstract}
\keywords{Left-to-right arithmetic, FPGA, adder tree, ultrasound beamforming, dynamic precision, energy efficiency.}

\section{Introduction}
\label{sec:intro}

Vision Transformers (ViTs)~\cite{dosovitskiy2020,touvron2021} have become widely used in computer vision because self-attention can model long-range spatial dependencies across image regions. This makes them effective for classification, detection, segmentation, and visual monitoring tasks in domains such as medical imaging, industrial quality control, smart-city sensing, autonomous-system perception, and infrastructure inspection~\cite{liao2022}. As these workloads move from cloud servers to cameras, embedded controllers, agricultural sensors, and inspection nodes, inference energy becomes a continuous operational cost rather than a one-time training cost. Prior work has therefore argued that the energy and carbon footprint of AI should be treated as a design constraint, not as an afterthought~\cite{strubell2019,patterson2022}.

This concern is particularly relevant for sustainability-oriented edge intelligence, where large numbers of low-power devices may monitor renewable-energy infrastructure, smart grids, and industrial systems. ViT-based models have already been explored for photovoltaic-panel inspection~\cite{ahmad2021}, photovoltaic defect segmentation~\cite{zhou2024pdet}, wind-power forecasting~\cite{huang2023windtransformer}, and electrical-load forecasting~\cite{lheureux2022load}. However, deploying ViTs on resource-limited FPGAs remains challenging because multi-head self-attention includes a softmax stage that requires exponentiation, summation, and division. While dot-product score computations map efficiently to FPGA DSP blocks, conventional softmax implementations often rely on CORDIC pipelines or BRAM-based look-up tables, increasing area, latency, and memory pressure. Base-2 approximations reduce this cost, but they can change the effective attention temperature and may require model-specific calibration to recover pre-trained accuracy.

To this end, we propose a BRAM-free attention-weighting unit that directly approximates the natural exponential used in softmax:
\begin{equation}
\tilde{p}_j =
\frac{\tilde{e}(u_j)}{\sum_k \tilde{e}(u_k)},
\qquad
u_j = s_j - \max_k s_k \leq 0 .
\end{equation}
Here, \(\tilde{e}(x)\) denotes a 16-segment uniform piecewise-linear approximation of \(e^x\) over the interval \([-8,0]\). Inputs below \(-8\) are saturated to the lower boundary, where the exponential contribution is already negligible. The approximation is implemented using 17 stored endpoint values, requiring only 272 bits of distributed LUTRAM for 16-bit entries, and therefore does not use BRAM.

Because the softmax scores are max-centered following the standard numerically stable softmax formulation \cite{blanchard2021accurately}, all exponential inputs satisfy \(u_j \leq 0\), making \([-8,0]\) a compact and hardware-friendly approximation range. Approximating the natural exponential directly also avoids the implicit logit rescaling introduced by a direct base-2 substitution, \(2^x = e^{x\ln 2}\) \cite{lin2021fq}. Therefore, the proposed unit preserves the pre-trained attention temperature without requiring base conversion or model-specific temperature calibration.

This paper makes the following contributions:
\begin{enumerate}
\item A BRAM-free 16-segment PWL approximation of the natural exponential \(e^x\) for ViT attention-weight generation, requiring no model-specific temperature calibration and preserving the pre-trained attention temperature.

\item A complete proof-of-concept 197-token attention-row arithmetic core on Xilinx Zynq-7020: 1{,}444~LUTs, 1{,}899~FFs, 77~DSPs, zero BRAM, with WNS~\(=+1.50\)~ns at \SI{100}{\mega\hertz}.

\item Post-route power analysis using Switching Activity Interchange Format (SAIF) switching traces shows 21~mW dynamic power and 124~mW total on-chip power, corresponding to \(1.66~\mu\mathrm{J}\) per row and 601.3~krows/s/W dynamic efficiency.

\item Hardware-accurate accuracy evaluation on Imagenette with ViT-S/16, ViT-B/16, and ViT-L/16 under an INT16 matched setting, showing a maximum top-1 accuracy change of \(0.20\%\) relative to the exact-softmax reference.

\item An illustrative 500-node deployment scenario quantifying a 43.3~MWh/year energy gap relative to a 10~W embedded-GPU reference, motivating BRAM-free FPGA attention hardware for sustainable edge-AI.
\end{enumerate}

\section{Background and Related Work}
\label{sec:background}

\subsection{Softmax Attention and FPGA Cost}

For one query row of one ViT attention head, let
\(q\in\mathbb{R}^{d_k}\) denote the query vector, and let
\(\{(k_j,v_j)\}_{j=1}^{N}\) denote the key-value pairs for the
\(N\) tokens in the sequence, where \(k_j\in\mathbb{R}^{d_k}\) is the key vector and \(v_j\in\mathbb{R}^{d_v}\) is the value vector for token \(j\). The scaled dot-product score \(s_j\) between the query \(q\) and key \(k_j\)
is computed as:
\begin{equation}
  s_j = \frac{q\cdot k_j}{\sqrt{d_k}},
  \label{eq:attention_score}
\end{equation}
where \(d_k\) denotes the dimensionality of the query and key vectors. In
fixed-point hardware, the scaling factor \(1/\sqrt{d_k}\) can be absorbed into the score representation.
The max-centered score \(u_j\) is then obtained by subtracting the row maximum from each score
\begin{equation}
  u_j = s_j - \max_{1\le l\le N} s_l,
  \label{eq:max_centered_score}
\end{equation}
where the maximum is taken over all \(N\) scores in the same attention row. This construction ensures that \(u_j \le 0\) for all \(j\), and that at least one entry satisfies \(u_j = 0\).

The attention weight \(p_j\) and output vector \(o\) are then computed as:
\begin{equation}
  p_j = \frac{e^{u_j}}{\sum_{l=1}^{N} e^{u_l}},
  \qquad
  o = \sum_{j=1}^{N} p_j v_j .
  \label{eq:attention_weight_output}
\end{equation}
Here, \(p_j\) is the normalized attention weight assigned to value vector \(v_j\), and \(o\in\mathbb{R}^{d_v}\) is the output vector for the current query row.

Several FPGA studies have proposed compact standalone softmax units for neural network inference. Wasef \etal~\cite{wasef2021multirate} and Mehra \etal~\cite{mehra2023empirical} reduce the cost of the softmax nonlinearity through hardware-oriented approximation and scheduling, but they do not include the surrounding attention-row operations required in ViT inference, such as score computation and value accumulation.

Transformer-oriented softmax accelerators are more directly related to this work. Li \etal~\cite{li2023approxsoftmax} approximate softmax using a base-2 Maclaurin formulation, reducing the cost of exponential evaluation for Transformer inference. Hirayae \etal~\cite{hirayae2025} combine a shift-based exponential approximation with the online softmax algorithm, reducing memory traffic by updating the normalization terms in a streaming manner. ViTA \cite{vita2023} instead uses an exact CORDIC-based softmax implementation, prioritizing numerical fidelity but at higher hardware cost. These works show that softmax is a central bottleneck in attention hardware, but they either rely on base-2 or shift-based approximations, online normalization, or exact CORDIC-style computation.

Other FPGA accelerators, including the Softmax/GELU unit~\cite{softmaxgelu2023}, Hyft~\cite{xia2024hyft}, and the parallel softmax architecture of Celep \etal~\cite{celep2025parallel}, emphasize throughput on larger FPGA platforms. Their resource assumptions differ from small SoC FPGAs, where BRAM, LUTs, and power are tightly constrained.

Related ASIC and integer-only Transformer designs also reduce nonlinear attention costs. I-BERT~\cite{ibert2021} replaces Transformer nonlinearities with integer polynomial approximations, ITA~\cite{islamoglu2023ita} uses integer and shift-based attention operations, Cross-Road~\cite{kim2024crossroad} explores base-2 and one-hot attention schemes, and Kawamura \etal~\cite{kawamura2025} apply PWL approximation to the full attention block with quantization-aware retraining.

Existing work establishes softmax approximation as an important direction for efficient Transformer hardware, but zero-BRAM attention weighting on constrained SoC FPGAs remains underexplored. Unlike base-2 or shift-based substitutions, which can alter the effective softmax temperature of a pre-trained model, this work directly approximates the natural exponential \(e^x\) using a small PWL table implemented in distributed LUT fabric. This enables BRAM-free attention weighting while preserving the pre-trained attention scale without post-training correction.

\section{Proposed Approximate Attention}
\label{sec:approx}

\subsection{Natural-Exponential PWL Weighting}
The proposed unit builds on the stable softmax formulation in Eq.~\ref{eq:attention_weight_output}. Since the scores are max-centered, the exponential input \(u_j\) satisfies \(u_j \le 0\), allowing the approximation domain to be restricted to \(u_j\in[-8,0]\). We replace only the exact exponential \(e^{u_j}\) with a piecewise-linear approximation \(\tilde{e}(u_j)\), while the normalization and value accumulation remain unchanged. The datapath consists of five main stages: score computation, PWL exponential weighting, denominator accumulation, value-weighted numerator accumulation, and final reciprocal scaling. This structure avoids CORDIC pipelines and BRAM-based exponential tables while preserving the natural-exponential softmax form used by the pre-trained model.

\subsection{Uniform Natural-Exponential PWL Approximation} 

We approximate the natural exponential \(e^x\) over the interval \(x\in[-8,0]\) using 16 uniform piecewise-linear segments. The segment boundaries \(b_i\) are defined as:
\begin{equation}
b_i = -8 + 0.5\,i,
\qquad i = 0,\ldots,16 .
\label{eq_bi}
\end{equation}

For an input \(x\in[b_i,b_{i+1})\), where \(i=0,\ldots,15\), the PWL
approximation \(\tilde{e}(x)\) is computed as:
\begin{equation}
\tilde{e}(x)
= y_i + \alpha\,(y_{i+1}-y_i),
\qquad
y_i = e^{b_i},
\label{eq:pwl}
\end{equation}

where \(y_i\) is the stored boundary value and \(\alpha\in[0,1)\) is the fractional offset of \(x\) within the segment. For the final boundary, \(x=0\) is assigned to the last segment endpoint.

\begin{figure}[!h]
\centering
\includegraphics[width=\columnwidth]{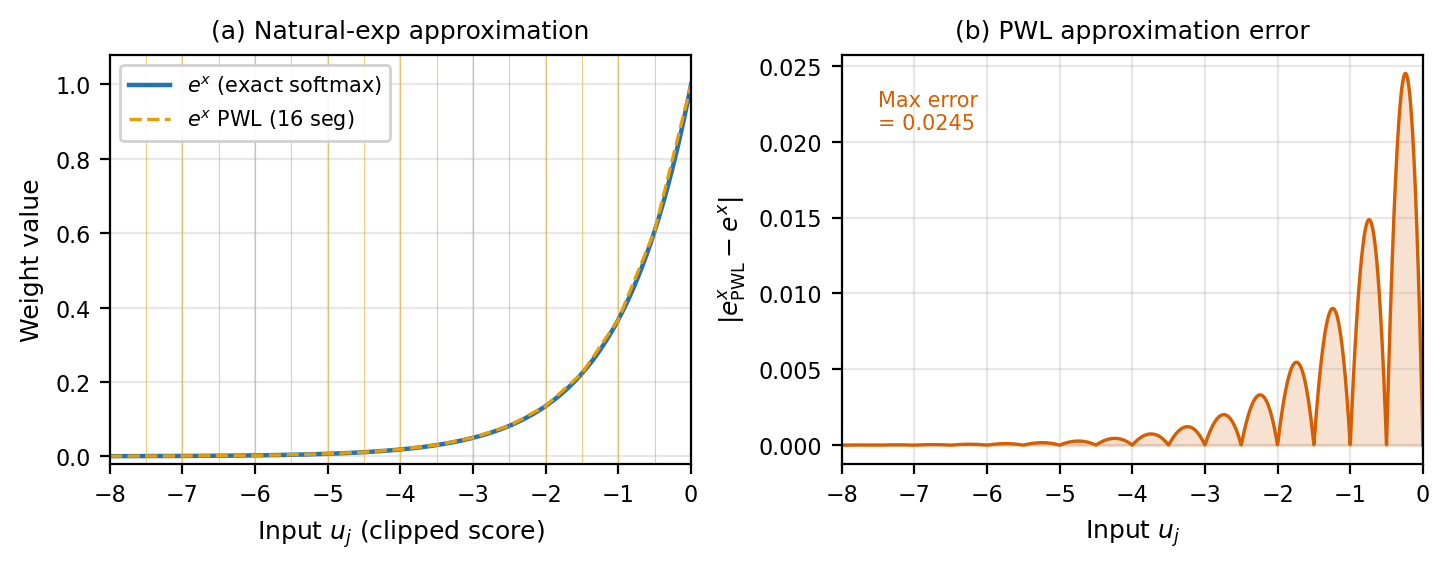}
\caption{(a)~Natural-exponential weighting kernel \(e^x\) over the clipped
stable-score domain \([-8,0]\), with 16-segment PWL boundaries overlaid.
(b)~Pointwise absolute error \(|\tilde{e}(x) - e^x|\); maximum error is
0.0245 in the segment closest to \(x=0\). The approximation is positive and monotone,
preserving softmax semantics without temperature recalibration.}
\label{fig:pwl}
\end{figure}

In the RTL implementation, the input \(x\) is represented in Q8.8 fixed-point format and is first clipped to the interval \([-8,0]\). The segment index and fractional offset are then extracted from fixed-point bit fields. The boundary values \(y_i\) are stored in a fixed-point fractional format, with the \(x=0\) endpoint handled by saturation if Q0.16 storage is used. The interpolation product is implemented in LUT logic, keeping the weight unit free of BRAM and DSP blocks.
Figure~\ref{fig:pwl} visualizes the approximation quality. The largest absolute error is 0.0245 and occurs in the segment closest to \(x=0\), where \(e^x\) has the greatest curvature over the approximation range. Within the non-clipped interval \([-8,0]\), the PWL approximation remains positive and strictly monotone, preserving the rank ordering of the unnormalized attention weights. Because the stored boundary values approximate \(e^x\) directly, no base conversion or per-model temperature scaling is required.

The 17 boundary values, each 16~bits, occupy 272~bits of distributed
LUTRAM, requiring only a small number of LUTs and zero BRAM.

\begin{figure}[!h]
\centering
\includegraphics[width=0.65\columnwidth]{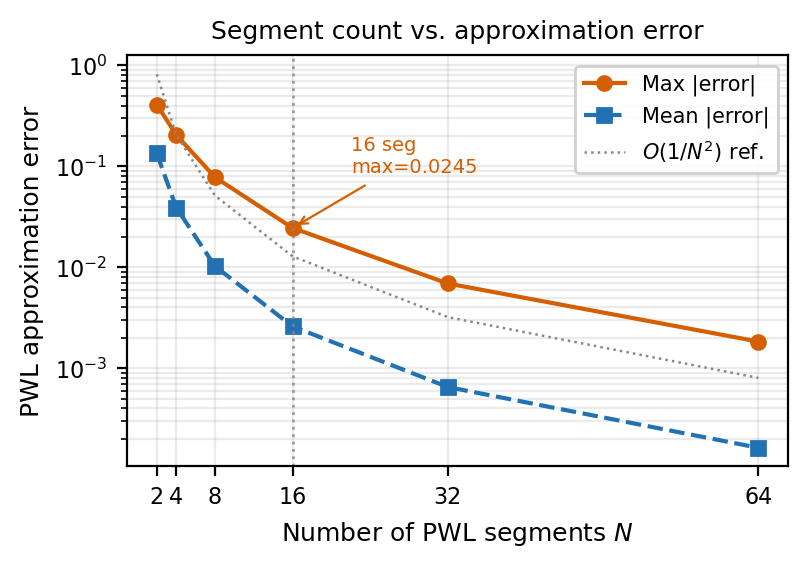}
\caption{Maximum and mean absolute PWL error vs.\ number of uniform
segments \(S\) over \([-8,0]\), with an \(O(S^{-2})\) reference.
At \(S=16\) (dashed), maximum error is 0.0245 and the coefficient table
fits entirely in 272~bits of distributed LUTRAM, requiring zero BRAM.}
\label{fig:segments}
\end{figure}

Figure~\ref{fig:segments} shows the trade-off between PWL approximation error and the number of uniform segments \(S\). For a fixed approximation interval, the maximum linear-interpolation error decreases as \(O(S^{-2})\). With \(S=16\), the maximum absolute error is 0.0245, and the 17-entry boundary table requires only 272~bits of distributed LUTRAM. Increasing to \(S=32\) reduces the maximum error to approximately 0.006, but nearly doubles the table size to 528~bits. We therefore use \(S=16\) as a practical operating point that balances approximation accuracy with LUTRAM footprint.

\section{Hardware Implementation}
\label{sec:hw}

\subsection{Core Architecture}

The design targets a Xilinx Zynq-7020 (xc7z020clg484-1) and implements one
complete attention row for \(N=197\) tokens, \(d_k=64\), \(d_v=64\),
matching a ViT class-token plus patch-token row.
The core uses a numerically stable two-pass schedule: \emph{Pass~1} computes
all dot-product scores \(s_j\) and tracks the row maximum; \emph{Pass~2}
replays each token, evaluates the PWL natural-exponential weight
\(\tilde{e}(u_j)\), and accumulates the numerator and denominator vectors.
The restoring vector divider then normalizes the output.

Figure~\ref{fig:arch} shows the five-module datapath. Five pipeline modules
implement this schedule (Table~\ref{tab:util}); their
post-route resources are detailed below:

\begin{figure}[!ht]
\centering
\includegraphics[width=1.0\columnwidth]{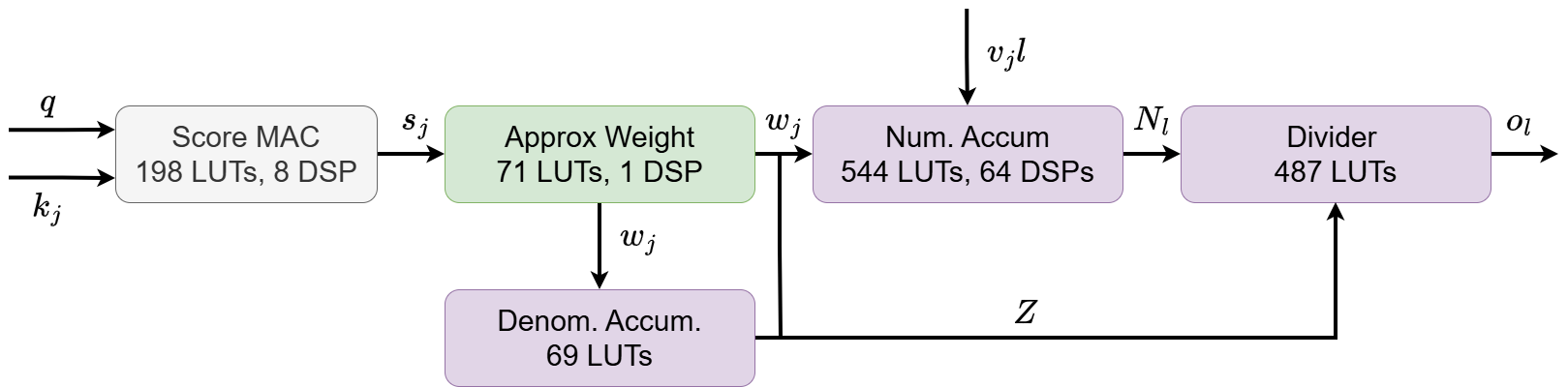}

\caption{Five-module attention-row datapath for the conventional DSP-MAC design.}
\label{fig:arch}
\end{figure}

The attention-row core is organized into five stages. The score path uses DSP48E1 MAC lanes to compute the scaled dot-product scores for \(d_k=64\). The PWL weight path clips \(u_j\) to \([-8,0]\), selects one of 16 segments, and evaluates \(\tilde{e}(u_j)\) using the 272-bit distributed LUTRAM table. The numerator and denominator paths accumulate \(\tilde{e}(u_j)v_j\) and \(\tilde{e}(u_j)\), respectively. Finally, a serialized 16-bit restoring divider normalizes the accumulated numerator over the 64 output dimensions. Per-module latency and resource usage are reported in Table~\ref{tab:util}.

Per-head latency:
\(T_\mathrm{head} = 197 {\times} 14 + 1{,}024 = 3{,}782\)~cycles, or \(37.8~\mu\mathrm{s}\), at
\SI{100}{\mega\hertz} (score-plus-weight pass).
Full row schedule including two passes and division:
\(T_\mathrm{row} = 7{,}920\)~cycles, or \(79.2~\mu\mathrm{s}\).

\subsection{Synthesis and Power Results}

The RTL is synthesized and implemented with Vivado~2024.1 targeting
\SI{100}{\mega\hertz}. Table~\ref{tab:util} reports post-route utilization.
Post-route power uses SAIF-driven activity from an RTL simulation of a
complete 197-token row; the Vivado report marks the estimate as high
confidence with \(\geq 95\%\) input and clock activity coverage.

\begin{table}[!h]
\centering
\caption{Module-Level Resource Utilization and SAIF Dynamic Power
(Zynq-7020, 197-token row, \(d_k=d_v=64\), 100~MHz, Vivado~2024.1)}
\label{tab:util}
\setlength{\tabcolsep}{20pt}
\footnotesize
\resizebox{\columnwidth}{!}{%
\begin{tabular}{lrrrc}
\toprule
Module & LUT & DSP & BRAM & Power (mW) \\
\midrule
Score MAC              & 198  &  8 & 0 & 11 \\
Num.\ Accumulator      & 544  & 64 & 0 & N/A \\
Natural PWL Weight     &  71  &  2 & \textbf{0} &  3 \\
Denom.\ Accumulator    &  69  &  0 & 0 &  1 \\
Restoring Divider      & 487  &  0 & 0 &  6 \\
\midrule
\textbf{Post-route total}$^*$ & \textbf{1{,}444} & \textbf{77} & \textbf{0} & \textbf{21 dyn. / 124 total} \\
Available (Zynq-7020)  & 53{,}200 & 220 & 140 & N/A \\
Utilization            & 2.71\% & 35.0\% & 0\% & N/A \\
\bottomrule
\multicolumn{5}{l}{\scriptsize $^*$Includes Vivado top-level glue/control overhead: +75 LUTs and +3 DSPs.}
\end{tabular}}
\end{table}

The design meets timing with WNS~\(=+1.50\)~ns at \SI{100}{\mega\hertz}.
The dominant DSP consumer is the Numerator Accumulator (64~DSP48E1s for
parallel value accumulation). The Natural PWL Weight unit contributes only
71~LUTs, 2~DSPs, and zero BRAM, confirming that natural-exponential
weight generation is the least resource-intensive stage.
Dynamic power is dominated by DSP switching in the score MAC and
numerator accumulator; the weight unit contributes approximately 3~mW.

\section{Evaluation}
\label{sec:eval}

\subsection{Approximation Accuracy} 
The 16-segment natural PWL table was evaluated using hardware-accurate software emulation of the RTL datapath. Table~\ref{tab:accuracy_int8} reports top-1 accuracy on the full \(3{,}550\)-image Imagenette~\cite{imagenette} validation split for ViT-B/16 and DeiT-S/16. Pre-trained weights are loaded from \texttt{timm}~\cite{timm2019} without modification. For the INT8 setting, \(Q\), \(K\), and \(V\) are fake-quantized using per-tensor symmetric quantization. Across both models, the approximate INT8 softmax remains within \(0.2\%\) of the exact INT8 softmax and within \(0.5\%\) of the FP32 exact-softmax reference. The cosine similarity to the exact INT8 baseline exceeds 0.988, indicating that the PWL exponential introduces only a small perturbation to the attention output.
\begin{table}[!h]
\centering
\caption{Approximation accuracy for INT8 evaluation on the \(3{,}550\)-image Imagenette validation split using 8-bit \(Q\), \(K\), and \(V\) quantization and a 16-segment natural-exponential PWL approximation.}

\label{tab:accuracy_int8}
\setlength{\tabcolsep}{20pt}
\footnotesize
\resizebox{\columnwidth}{!}{%
\begin{tabular}{llccc}
\toprule
Model & Mode & Top-1 Acc. & Cos.\ Sim. & Top-1 Agr. \\
\midrule
\multirow{4}{*}{ViT-B/16}
  & Exact FP32          & 84.0\% & 1.000 & 100.0\% \\
  & Exact, sim.\ INT8   & 84.6\% & 0.988 &  97.6\% \\
  & Approx FP32         & 83.9\% & 0.998 &  99.2\% \\
  & \textbf{Approx INT8} & \textbf{84.4\%} & \textbf{0.988} & \textbf{97.4\%} \\
\midrule
\multirow{4}{*}{DeiT-S/16}
  & Exact FP32          & 84.6\% & 1.000 & 100.0\% \\
  & Exact, sim.\ INT8   & 84.9\% & 0.996 &  98.8\% \\
  & Approx FP32         & 84.7\% & 0.999 &  99.5\% \\
  & \textbf{Approx INT8} & \textbf{84.6\%} & \textbf{0.995} & \textbf{98.7\%} \\
\bottomrule
\end{tabular}}
\end{table}

For the INT16 comparison, we follow the matched protocol of Li \etal~\cite{li2023approxsoftmax}, using INT16 quantization with ViT-S/16, ViT-B/16, and ViT-L/16. Table~\ref{tab:accuracy_int16} reports the top-1 accuracy difference relative to the FP32 exact-softmax reference. Across all three models, the largest degradation is \(0.20\%\) for ViT-S/16, while ViT-B/16 and ViT-L/16 remain within \(0.05\%\) of the reference. These results indicate that the proposed natural PWL approximation preserves accuracy under the INT16 matched setting.

\begin{table}[!ht]
\centering
\caption{Top-1 accuracy under the INT16 matched protocol on the \(3{,}550\)-image Imagenette validation split without fine-tuning.}
\label{tab:accuracy_int16}
\setlength{\tabcolsep}{20pt}
\footnotesize
\resizebox{\columnwidth}{!}{%
\begin{tabular}{lrrrr}
\toprule
Model & FP32 exact & INT16 exact & RTL-PWL INT16 & \(\Delta\) top-1 \\
\midrule
ViT-S/16 & 74.99\% & 74.96\% & 74.79\% & \(-0.20\%\) \\
ViT-B/16 & 83.97\% & 83.97\% & 83.94\% & \(-0.03\%\) \\
ViT-L/16 & 84.54\% & 84.54\% & 84.59\% & \(+0.05\%\) \\
\bottomrule
\end{tabular}}
\end{table}

Li \etal~\cite{li2023approxsoftmax} report top-1 accuracy differences of \(0.25\%\), \(0.34\%\), and \(0.39\%\) for ViT-S/16, ViT-B/16, and ViT-L/16 on ImageNet-1K using a base-2 Maclaurin softmax unit under INT16 quantization. Under the same precision and model-family setting, our natural PWL softmax gives top-1 accuracy differences of \(0.20\%\), \(0.03\%\), and \(-0.05\%\) on Imagenette. Because the datasets and checkpoints differ, these numbers should be interpreted as a robustness check under a matched precision setting rather than as a direct accuracy ranking.

\subsection{FPGA Implementation Results}

\begin{figure}[!ht]
\centering
\includegraphics[width=\columnwidth]{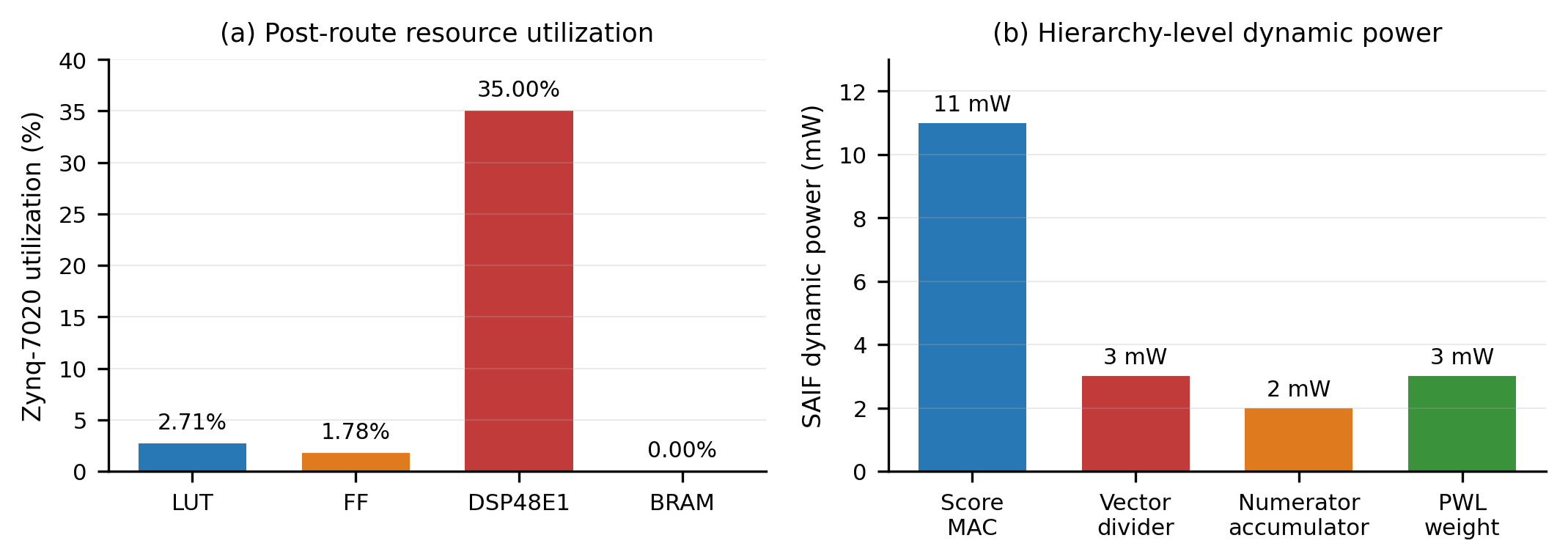}
\caption{Post-route implementation results on Zynq-7020.
(a)~Resource utilization by module relative to device capacity
(LUT: 2.71\%, FF: 1.78\%, DSP: 35.0\%, BRAM: 0\%).
(b)~Hierarchy-level SAIF-driven dynamic power.
The Natural PWL Weight unit (3~mW) is not the dominant power consumer.}
\label{fig:results}
\end{figure}

\begin{table*}[t]
\centering
\caption{Comparison with FPGA softmax and attention hardware.
Standalone softmax blocks exclude score generation and accumulation.}
\label{tab:hwcomp}

\footnotesize
\setlength{\tabcolsep}{3pt}
\renewcommand{\arraystretch}{1.15}

\begin{tabularx}{\textwidth}{@{}Z{2.7cm}Z{3.0cm}Z{3.2cm}Z{2.6cm}Y@{}}
\toprule
Reference 
& FPGA / Design scope 
& FPGA resources 
& Performance 
& Power / accuracy impact \\
\midrule

Wasef \& Rafla~\cite{wasef2021multirate}
& ZED-board; standalone softmax
& 788 LUT, 577 FF, 5 BRAM, 2 DSP
& 150~MHz; 2 outputs/cycle
& 35~mW; RMSE $3.2{\times}10^{-7}$ \\

Koca et~al.~\cite{koca2023hardware}
& ZCU102; self-attention softmax, $n=8$
& 909 LUT, 333 FF, 0 DSP
& 476~MHz; 10.5~ns
& Power NR; $<1$\% BERT accuracy loss \\

Mehra et~al.~\cite{mehra2023empirical}
& Zynq; standalone CORDIC softmax
& 427 LUT, 0.5 BRAM
& 685~MHz; 1 output/cycle
& Power NR; $<2$\% accuracy loss \\

Li et~al.~\cite{li2023approxsoftmax}
& ZCU102; approximate Transformer softmax
& 1,872 LUT, 772 FF, 0 BRAM
& 133~MHz; 1.06~Gbps
& 49.56~mW; ViT accuracy change 0.25/0.34/0.39\% \\

Hirayae et~al.~\cite{hirayae2025}
& KV260, UltraScale+; shift-exp with online processing
& 1,319 LUT, 1 BRAM
& NR
& Power NR; $<0.2$\% accuracy loss \\

ViTA~\cite{vita2023}
& Zynq-7020; exact CORDIC attention head
& NR LUT, BRAM used
& NR
& 880~mW total power \\

Celep et~al.~\cite{celep2025parallel}
& ZCU104; parallel Transformer softmax
& 27,746--48,767 LUT, 8--32 DSP
& 200~MHz; 10--43~$\mu$s
& 1.56--1.64~W; 1.2\% deviation \\

Aboagye et~al.~\cite{aboagye2026generalizable}
& Zybo Z020; low-precision ViT softmax
& 16,728 LUT, 112 BRAM
& 70.4~MHz
& 0.4~W; 0.15~dB PSNR loss \\

\midrule

\textbf{This work}
& \textbf{Zynq-7020; complete 197-token attention row}
& \textbf{1,444 LUT, 1,899 FF, 77 DSP, 0 BRAM}
& \textbf{100~MHz; 79.2~$\mu$s/row}
& \textbf{21~mW dynamic; 124~mW total; $\leq 0.20$\% accuracy change} \\

\bottomrule
\end{tabularx}


\end{table*}

Figure~\ref{fig:results} summarizes the post-route resource and power breakdown. On the Zynq-7020, the core uses only \(2.71\%\) of LUTs, \(1.78\%\) of flip-flops, \(35.0\%\) of DSP48E1 blocks, and zero BRAM. The measured dynamic power is \(21~\mathrm{mW}\), lower than the \(35~\mathrm{mW}\) standalone softmax result reported by Wasef and Rafla~\cite{wasef2021multirate}. For a 197-token attention row at \(100~\mathrm{MHz}\), the RTL-simulated row latency is (7{,}920) cycles, corresponding to \(79.2~\mu\mathrm{s}\) and a single-core row rate of \(12.63~\mathrm{krows/s}\). This gives a dynamic throughput efficiency of \(601.3~\mathrm{krows/s/W}\), with \(1.66~\mu\mathrm{J}\) dynamic energy per row and \(8.44~\mathrm{nJ}\) per token. When total on-chip power is considered, the throughput efficiency is \(101.8~\mathrm{krows/s/W}\), corresponding to \(9.82~\mu\mathrm{J}\) per row. These figures exclude sensor input, external memory, DMA, and the remaining Transformer pipeline.



\subsection{Hardware Comparison}
\label{sec:compare}

Table~\ref{tab:hwcomp} compares the proposed design with prior exponentiation, softmax, and attention-weighting hardware. To keep the comparison boundary clear, we report both the Natural PWL Weight block and the complete attention-row core. The weight block uses only \(71\) LUTs and zero BRAM, which is \(26\times\) fewer LUTs than Li \etal~\cite{li2023approxsoftmax} and \(19\times\) fewer than Hirayae \etal~\cite{hirayae2025}. The full attention-row core uses \(1{,}444\) LUTs because it also includes score computation and value accumulation, which are outside the unit-level boundaries of~\cite{li2023approxsoftmax,hirayae2025}. Overall, the proposed design combines zero BRAM usage, \(21~\mathrm{mW}\) dynamic power, and at most \(0.20\%\) accuracy change under the matched INT16 evaluation, while targeting the resource-constrained Zynq-7020 rather than a larger UltraScale+ device.

\subsection{Illustrative Deployment Energy Scenario} Table~\ref{tab:carbon} scales the measured arithmetic-kernel power to an illustrative network deployment. We consider \(500\) continuously active monitoring nodes, such as distributed photovoltaic-panel inspection or wind-farm monitoring networks~\cite{huang2023windtransformer,zhou2024pdet}. The measured FPGA arithmetic core is compared with a nominal \(10~\mathrm{W}\) embedded-GPU reference module. This estimate is not a full-system carbon assessment; it excludes sensors, external memory, communication, host control, and the remaining Transformer pipeline.

\renewcommand{\arraystretch}{1.5}
\begin{table}[!ht]
\centering
\caption{Illustrative Annual Energy for 500 Continuously-Active Monitoring Nodes}
\label{tab:carbon}
\setlength{\tabcolsep}{22pt}
\footnotesize
\resizebox{\columnwidth}{!}{%
\begin{tabular}{lrrrr}
\toprule
Platform & Per node (W) & Network (kW) & Annual (MWh) & CO\(_2\)e (t/yr) \\
\midrule
10~W reference (embedded GPU) & 10.0 & 5.000 & 43.80 & 17.96 \\
\textbf{This work: FPGA core} & \textbf{0.124} & \textbf{0.062} & \textbf{0.54} & \textbf{0.22} \\
\midrule
Illustrative gap & 9.876 & 4.938 & \textbf{43.26} & \textbf{17.74} \\
\bottomrule
\end{tabular}}
\end{table}

For the FPGA estimate, we use the conservative total on-chip power of
\(124~\mathrm{mW}\), rather than only the \(21~\mathrm{mW}\) dynamic
component. At \(500\) nodes, the resulting energy gap is
\(43.26~\mathrm{MWh/year}\), corresponding to approximately
\(17.74\) tonnes CO\(_2\)e/year using an emissions factor of
\(0.41~\mathrm{kg}\) CO\(_2\)/kWh~\cite{patterson2022}. At \(50\%\) and \(10\%\) duty cycles, the gaps scale to \(21.6\) and
\(4.3~\mathrm{MWh/year}\), respectively. These values are arithmetic-kernel estimates, not complete-system energy measurements. They exclude Q/K/V buffers, projections, memory interfaces, and the remaining Transformer layers. The purpose of this scenario is therefore to quantify potential power-budget headroom and motivate full-system board-level measurement as future work.



\section{Conclusion}
\label{sec:conclusion}

We presented a BRAM-free approximate attention-weighting unit for FPGA-based
Vision Transformer inference that directly approximates the natural
exponential \(e^x\) with a 16-segment uniform PWL function stored in 272~bits
of distributed LUTRAM.
Unlike base-2 alternatives, the proposed design preserves pre-trained
softmax semantics without model-specific temperature calibration.
Implemented on Xilinx Zynq-7020 with SAIF-driven power analysis, the
complete 197-token attention-row core occupies 1{,}444~LUTs, 77~DSPs,
and zero BRAM, with 21~mW SAIF-measured dynamic power and 124~mW
total on-chip power at \SI{100}{\mega\hertz}, corresponding to
\(1.66~\mu\mathrm{J}\) dynamic energy per row (601.3~krows/s/W).
Hardware-accurate accuracy evaluation shows the natural PWL approximation
stays within \(0.20\%\) of FP32 exact softmax on ViT-S/B/L
(INT16, Imagenette), remaining comparable to prior FPGA softmax units
without calibration.
An illustrative 500-node sustainability scenario quantifies a
43.26~MWh/year energy gap relative to a 10~W embedded-GPU reference,
supporting the potential of zero-BRAM FPGA attention hardware as a
building block for sustainable edge-AI monitoring.
Full-system implementation, board-level power measurement, and
application-dataset validation remain as future work.
\section*{Acknowledgement}
This work was supported by the German Research Foundation
(Deutsche Forschungsgemeinschaft, DFG) under project number 573796083.

\bibliographystyle{IEEEtran}
\bibliography{references_v3}

\end{document}